  \def\selectedoptions{final}

\documentclass[
   \selectedoptions
  ]
  {aipproc}

\layoutstyle{6x9}

\SetInternalRegister\hbadness{8000} 

\newcommand\doingARLO[2][]{%
  \ifx\mmref\undefined #1\else #2\fi
}
\newcommand\be{\begin{equation}}
\newcommand\ee{\end{equation}}
\newcommand\tpo{$^3{\rm P}_0$\ }
\newcommand\dego{\rm o}
\newcommand\bwp{$b_1 \to \omega \pi$\ }
\newcommand\arp{$a_1 \to \rho \pi$\ }
\newcommand\qqbar{$q\bar q$\ }

\begin{document}

\title 
{Strong Decays: Past, Present and Future}

\keywords{}

\author{T.Barnes}{
  address={Department of Physics, University of Tennessee, Knoxville, TN 37996, USA},
  address={Physics Division, Oak Ridge National Laboratory, Oak Ridge, TN 37831, USA},
  email={tbarnes@utk.edu},
}

\begin{abstract}
In this talk I review the history of models of strong decays, from the
original model through applications to charmonium, light and charmed mesons,
glueballs and hybrids.
Our current rather limited understanding of the QCD mechanism of strong decays
is stressed. Regarding current and future applications of strong decay models,
we note that in certain channels the very strong
coupling predicted between $q\bar q$ basis states and the two-meson continuum
may lead to strongly mixed states and perhaps molecular two-meson bound states.
The relevance to the D$_{sJ}^*(2317)$ is discussed.
\end{abstract}


\maketitle

\section*{Historical Introduction}
                                                                                
\subsection*{Origins of the \tpo model}

\subsubsection*{Micu}

The earliest reference in which the currently widely accepted microscopic
physics of strong decays appears is due to Micu \cite{Micu69}, in the paper
"Decay Rates of Meson Resonances in a Quark Model". Micu was concerned
with understanding the light P-wave mesons in the quark model, especially
their widths. In the quark model, describing decays required the production
of a $q\bar q$ pair; in lieu of a microscopic model of interactions between
quarks she made the plausible assumption that the pair was produced with
vacuum ($0^{++}$) quantum numbers, therefore in a \tpo state. No explicit quark
model wavefunctions were assumed, so the implicit overlap integrals were
described by two free parameters, taken from data.
Micu applied this simple
model to approximately 30 known light meson decays, and
found that it was reasonably successful in explaining the observed
partial widths. She concluded that
``... this decay model proves once more the surprising viability of
the quark model."
                                                                                
\subsubsection*{The ORSAY group}

This work was follows by a series of decay model calculations by the
ORSAY group (LeYaouanc {\it et al.}), who introduced explicit nonrelativistic
quark model wavefunctions in the calculations of the decay amplitudes and
a fundamental pair production amplitude~$\gamma$. This cast the \tpo decay
model in essentially the form in which it is used today.
                                                                                     
\vskip 0.2cm
In their initial 1973 paper \cite{LeY73} the ORSAY group
set up the formalism of \tpo decays in the quark model and applied it to
light baryon and meson decays and couplings. The importance of the
D/S amplitude ratios in the decays \bwp and \arp as crucial tests of the
assumed \tpo quantum numbers of the \qqbar pair was first stressed in this
paper; this early and striking success of the \tpo model was strong evidence
in favor of this model of strong decays. This paper also notes that the
model predicts strong three-meson effective couplings and form factors, which
is an application that has not yet been widely exploited, but is of great
importance for the currently fashionable meson effective lagrangians and meson
exchange models of hadronic reactions.

\vskip 0.2cm
This introductory paper was followed by detailed studies of
baryon decays \cite{LeY74,LeY75}, which considered
$\approx 100$ baryon decay amplitudes. Finally, with the discovery of
charmonium and states above open-charm threshold, the open-charm decays
of charmonia were considered \cite{LeY77}, and it was noted that the
$\psi(4040)$ (elsewhere suggested as a ${\rm D}^*{\rm D}^*$ molecule candidate
due to its anomalous strong branching fractions) could be accepted as a
conventional $3{}^3$S$_1$ $c\bar c$ state, since the nodes of the
radial wavefunction could plausibly weaken the
${\rm D}{\rm D}$ and ${\rm D}{\rm D}^*$ modes.
                                      
\vskip 0.2cm
It is straightforward to give Feynman rules for this \tpo model, since
as noted by Ackleh {\it et al.} \cite{Ack96} it corresponds
to the nonrelativistic limit of a $\bar \psi \psi$ decay interaction. 
The two diagrams that describe $q\bar q$ decays are shown in Fig.1 below.    

\vskip 0.4cm
\begin{figure}[ht]
  \resizebox{18pc}{!}{\includegraphics{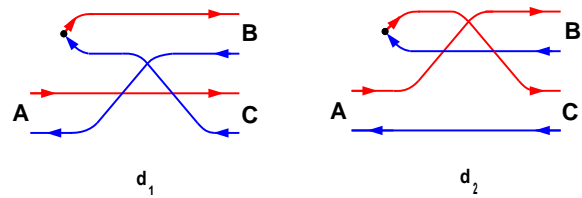}}
\caption{The two independent meson decay diagrams in the \tpo model.}
\end{figure}
\subsection*{The Cornell charmonium decay model} 

The discovery of charmonium motivated the next important development
in decay models. This was the introduction of a new microscopic model
for the QCD mechanism underlying strong decays, as well as studies of
coupled-channels effects, in a
series of papers by the Cornell group \cite{Eic76,Eic78,Eic80}.
                                       
\vskip 0.2cm
Eichten {\it et al.} assumed that strong decays were driven by
$q\bar q$ pair production from the linear confining interaction.
This implied that the size of the phenomenological \qqbar
pair production amplitude $\gamma$ in the \tpo model was proportional
to the \qqbar string tension, to the extent that these models could be
compared.

\begin{figure}[ht]
  \resizebox{11pc}{!}{\includegraphics{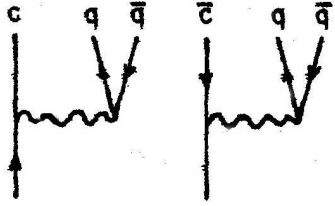}}
\caption{The Cornell decay model: 
pair production from linear vector confinement.}
\end{figure}
                                                                                \vskip 0.2cm
A crucial difference between the Cornell model and most other decay
and confinement models was their assumption of
{\it timelike vector confinement}. This was subsequently rejected
due to its inaccuracy in describing the splittings of the P-wave
$\chi_J$ multiplet; vector confinement gives a positive spin-orbit term that
does not agree with the data, whereas scalar confinement has negative
spin-orbit that leads to a good description of the $\chi_J$ masses. 
This issue
is certainly not settled, and recent theoretical work by Swanson and
Szczepaniak again argues in favor of vector confinement models \cite{Swa98}.
                                                                                     
\vskip 0.2cm
The Cornell group did not give individual decay amplitudes and partial widths
for resonances above open-charm threshold, but instead used a 
``resolvant" formalism to infer their combined 
contribution to $R$. This gave results that are quite similar
to the current experimental $R$, and (even if this model of confinement proves
incorrect) shows that this decay model merits further
study.
                                                                                     
\vskip 0.2cm
Finally, the very topical issue of the level of mixing between ``bare"
$c\bar c$ basis states and the two-meson continuum was also considered
by the Cornell group. They found that this is a significant effect,
with (in their final version of this model \cite{Eic80})
the 1P $\chi_J$ states being $\approx $ 10\% meson-meson and
the 2S $\psi'$ $\approx $ 20\% meson-meson. They warn however that their
two-meson intermediate states are truncated at ${\rm D}{\rm D}^*$, so the
actual two-meson component may be much larger.
                                                                                     
\section*{Modern ``surveys" of decay modes}

\subsection*{$\xi(2230)$ brou ha ha}

Most of the detailed ``survey" studies of strong decays that have appeared
since 1980 have assumed the \tpo model, either as originally formulated
or in one of several variants that modify the phase space, kinematics
or the spatial dependence of the pair production amplitude.
These survey papers typically attempt to consider as many open-flavor
decay modes as possible, due to the ``brou ha ha" associated with the
question of the $\xi(2230)$.

\vskip 0.2cm
The $\xi(2230)$ was originally reported
in $J/\psi$ radiative decays by MarkIII at SLAC \cite{Bal86}
in KK final states (apparently both charged and neutral kaons). The state
appeared remarkably narrow, hence it was considered a possible glueball
candidate. However there were problems with confirmation of the $\xi(2230)$;
DM2 for example did not see it, and they had slightly
better statistics than Mark III. For those with long memories,
the reports of the $\Theta(1542)$ at this meeting are disturbingly
reminiscent of the $\xi(2230)$.

\vskip 0.2cm
Although the narrow width of the $\xi(2230)$ made it a plausible tensor
glueball candidate, assuming that it did exist, this exciting possibility
was dampened by a ``spoiler" paper by Godfrey {\it et al.} \cite{God84},
who noted that
L=3 $s\bar s$ states were expected near this mass, and they should
``accidentally" be quite narrow (due to limited phase space and
centrifical barriers), provided that their decays were dominated
by decays to two S-wave mesons (KK, KK$^*$ and K$^*$K$^*$). The
$^3$F$_2$ $s\bar s$ assignment was preferred because $gg$ intermediate
states in $J/\psi$ radiative decays were predicted to 
populate J$^{\rm PC} = 0^{\pm +}$ and $2^{++}$ final states preferentially.

\vskip 0.2cm
This {\it a priori} plausible assumption of dominant decays to
S-wave meson pairs was subsequently tested by Godfrey {\it et al.}
\cite{Blu96}, who carried out a more complete set of decay calculations.
They were surprised to find that the decays of a $^3$F$_2$ $s\bar s$ meson 
were {\it not} dominated by S-wave meson pair final states. Instead the
dominant mode was the S+P final state KK$_1(1270)$ by a considerable margin, 
so this $s\bar s$ state was instead predicted to be rather broad, once all the
higher modes were included. Thus the $\xi(2230)$ could no longer be claimed 
to be an accidentally narrow $^3$F$_2$ $s\bar s$ state.

\vskip 0.2cm
This specific result has broad implications. Evidently one should be
cautious in using strong decay calculations to estimate total widths;
calculations of {\it all} open-flavor 
decay amplitudes and partial widths are prudent. The assumption of dominance by
a specific set of low-lying S-wave final mesons may well be inaccurate. 
For this reason, several recent papers on strong decays of light mesons 
have given results for all allowed open-flavor modes, and this approach should
probably be followed in future studies of baryon and heavy meson strong decays
as well. 

\subsection*{Recent decay surveys}

The past two decades have seen a series of detailed calculations of strong
decay amplitudes and partial widths using the \tpo model or a variant.
For mesons the best known is the encyclopaedic work by Godfrey and Isgur
\cite{God85}, and for baryons the corresponding paper by 
Capstick and Isgur \cite{Cap86}. These studies have been followed by 
more complete studies of decay modes. The paper of
Barnes, Close, Page and Swanson \cite{Bar97} evaluates {\it all} open
two-body modes of all nonstrange mesons expected in the quark model to
2.1~GeV; this was recently extended to all open modes of 
strange mesons in the same multiplets by Barnes, Black and Page   
\cite{Bar03a}. Calculations of nonstrange baryon decays were extended to include 
modes with vector mesons by Capstick and Roberts \cite{Cap94}; an extensive
review of baryon strong decays was recently published by the same authors
\cite{Cap00}.

\subsection*{New results for decays of strange mesons}

As this is a conference that is largely devoted to the physics of mesons,
it may be of interest to review some of the recent theoretical results
on strange meson decays \cite{Bar03a}. First, regarding the 
axial vectors, from the very broad width predicted for a $1^{++}$ state at the
mass of the $f_2'(1525)$ (ca. 400~MeV) it is clear that there is no ``hidden"
axial vector state with the same mass and width as the $f_2'$. 
The near-threshold $f_1(1420)$ does appear consistent with the predicted
width, although the nearby KK$^*$ threshold will modify the shape of this 
resonance. Second, one often hears the speculation that $s\bar s$ mesons
might be rather narrow, and hence offer attractive experimental targets.
In our survey the five narrowest $s\bar s$ states we found without widely
accepted experimental candidates were 

\vskip 0.2cm
\noindent
$^1$D$_2$ $\eta_2 (1850)$,  $\Gamma = 129$~MeV, 
dominant mode KK$^*$

\vskip 0.2cm
\noindent
$^3$F$_4$ $f_4 (2200)$,  $\Gamma = 156$~MeV, dominant modes 
KK, KK$^*$, K$^*$K$^*$

\vskip 0.2cm
\noindent
3$^1$S$_0$ $\eta (1950)$,  $\Gamma = 175$~MeV, dominant modes KK$^*$, K$^*$K$^*$

\vskip 0.2cm
\noindent
2$^1$P$_1$ $h_1 (1850)$,  $\Gamma = 193$~MeV, dominant modes 
KK$^*$, K$^*$K$^*$, 
$\eta \phi$

\vskip 0.2cm
\noindent
$^3$D$_2$ $\phi_2 (1850)$,  $\Gamma = 214$~MeV, dominant modes KK$^*$, 
$\eta \phi$

\vskip 0.2cm
Evidently these are only moderately narrow states. As one might expect the
final states KK, KK$^*$ and K$^*$K$^*$ are important for $s\bar s$. 
The mode $\eta \phi$
is much more attractive, however, as we do not expect $n\bar n$ ($n=u,d$)
mesons to couple significantly to this channel. This mode is in effect an
``$s\bar s$ filter", and should be much more attractive for identifying 
$s\bar s$ states than the open-strange modes involving kaons.

\vskip 0.2cm
Turning to open-strange mesons, we find that the ``strangest" state in the 
known strange meson spectrum is the K$^*(1414)$. First, the mass of 
this state appears much too light for a 2S radial vector kaon, given 
the nonstrange
candidates $\rho(1465)$ and $\omega(1419)$. It is also very surprising that 
it would have a lower mass than the radial pseudoscalar kaon K$(1460)$.
The decays of the K$^*(1414)$ are also a problem; in the \tpo model the dominant
mode is predicted to be $\pi $K, with a branching fraction of about $30\% $.
Experimentally, the branching fraction observed by LASS was only 
$6\%$. These discrepancies in mass and decays suggest a problem with 
a simple 2S radial K$^*$ assignment. One exciting possibility is that we
may be seeing the effect of mixing with exotic vector hybrid states; since 
there is no C-parity in kaons, the $1^{--}$ and $1^{-+}$ basis states mix,
giving an overpopulation of vector states and a different mass matrix for
$1^-$ kaons than for $1^{--}$ $\rho$ and $\omega$ excited vectors. A simple
comparison of strange and nonstrange excited vectors may therefore show the
presence of the additional $1^{-+}$ hybrid basis states, and with significant 
hybrid mixing we would expect the excited kaons to have rather different masses 
and decay amplitudes than their nonstrange partners.   

\vskip 0.2cm
There are also very interesting issues regarding the singlet-triplet mixing
angles of excited kaons (analogous to the K$_1(1273)$-K$_1(1402)$ 
mixing angle), which can be determined from decay amplitudes of these excited
states. Another interesting aspect of kaon strong decays is the relative
strength of $\eta$ and $\eta'$ modes, which depends on the angular quantum 
numbers in a complicated manner due to an interference between the $n\bar n$
and $s\bar s$ components of the $\eta$ and $\eta'$ \cite{Bar03a}. These 
selection rules also have applications to B decays involving an 
$\eta$ or $\eta'$ \cite{Lip00}. 

\section*{Strong decays of exotica}

\subsection*{Hybrids}

The most interesting and influential predictions of meson decay models
in recent years have been the predictions for the decay modes of exotica,
specifically hybrid (excited glue) mesons. In the flux tube model of Isgur 
and Paton, hybrids are treated as states of quark, antiquark and flux-tube,
in which the flux-tube is spatially excited about the quark-antiquark axis.
When this simple quantum mechanical picture is combined with a \tpo model for
pair production, which is assumed to take place along the path of the 
flux tube,
one obtains a simple, intuitive picture of the open-flavor decay amplitudes of
gluonic hybrid mesons.

\vskip 0.2cm
Isgur, Paton and Kokoski \cite{Isg85} found that this simple picture gave a
plausible explanation as to why the predicted 
rich spectrum of hybrids had not been 
clearly identified; the preferred decays to conventional two-meson final 
states showed a strong preference for the so-called ``S+P modes", in which 
one of the final mesons had an internal orbital excitation. Since the P-wave
mesons have secondary decays, a search for hybrids would require a study of 
complicated multimeson final states, which had not previously been considered 
systematically.  In addition the numerical scales of the strong widths 
of hybrids were in many cases rather large, so that the hybrids would be 
difficult to identify. Isgur {\it et al.} cited the I=1 exotic with 
J$^{PC}=1^{-+}$ as a case of special interest, since the total width of this
state was found to be rather narrow, $\Gamma \approx 150$~MeV, for an initial 
hybrid mass of 1.9~GeV. The preferred decay modes of this hybrid in the 
flux tube decay model were $b_1\pi $ and $f_1\pi $; this observation has 
stimulated several experimental studies of these rather complicated 
final states.

\vskip 0.2cm
Isgur {\it et al.} only considered the decays of the J$^{PC}$-exotic hybrids.
These flux tube hybrid  
decay calculations were extended to nonexotic hybrids by Close and Page 
\cite{Clo95}, who found that some very interesting and relatively narrow 
nonexotics should appear in the spectrum as an overpopulation relative 
to the naive quark model, provided that the flux tube picture of hybrids and
their decays is reasonably accurate. Two notable cases are an extra $\omega$
(predicted to be only 100~MeV wide, and to decay mainly to the KK$_1$ channels)
and an extra $\pi_2$. This nonexotic hybrid $\pi_2$ is predicted to have 
a very characteristic $b_1\pi $ mode; this mode is forbidden 
to the conventional $q\bar q$ quark model $\pi_2$ (presumably the 
$\pi_2(1670)$), because it is an 
$
(S_{q\bar q} = 0)          
\to
(S_{q\bar q} = 0)          
+
(S_{q\bar q} = 0)          
$
transition. These are forbidden for $(q\bar q) \to (q\bar q)+(q\bar q)$
in the \tpo model as well as in OGE pair production and linear scalar 
pair production models. In contrast they are allowed for decays of 
flux-tube hybrids, since the $\pi_2$ hybrid actually has the quarks in an
$S_{q\bar q} = 1$ configuration. The very strong experimental limit of
the $b_1\pi $ mode of the $\pi_2(1670)$ reported by VES is implicitly 
a constraint on the size of the hybrid component of the $\pi_2(1670)$ 
state vector.

\vskip 0.2cm
As a final topic in hybrid decays,
there have been very interesting recent results from LGT on the 
{\it closed-flavor} decays of heavy hybrids. The UKQCD collaboration
\cite{McN02} has found that some of these modes, notably to a $\chi$ state
and a scalar (presumably an effectively $\pi\pi$) are remarkably large.
This result is very surprising in view of the weakness of 
the known closed-flavor $c\bar c$ and
$b\bar b$ dipion decays, but if correct suggests very attractive
modes for heavy hybrid searches, such as 
$(\gamma\; {\rm J}/\psi) + (\pi\pi)_S$ for charmonium hybrids. 

\subsection*{Glueballs}

The search for glueballs is a central component of the more general 
search for exotica in hadron spectroscopy. The well known LGT study 
of the glueball spectrum by Morningstar and Peardon \cite{Mor99} shows that
the currently experimentally accessible glueballs have non-exotic quantum 
numbers, since the lightest predicted glueball exotic is a $2^{+-}$ state
at a very high mass of $\sim 4$~GeV. At the presently accessible masses of
up to ca. 2.5~GeV just three glueballs are anticipated, a $0^{++}$ near
1.6~GeV and a $0^{-+}$ and $2^{++}$ near 2.3~GeV. The scalar channel
is most interesting at present because we have two experimental candidates
for this lightest glueball, the $f_0(1500)$ and $f_0(1710)$.

\vskip 0.2cm
Although width and decay mode predictions for glueballs are obviously of
paramount importance, there has been little work in this area. The naive
expectation that glueballs should have flavor-blind decays is clearly strongly
violated by both experimental candidates; the $f_0(1500)$ shows 
a strong preference for $\pi\pi$ modes over KK, and the $f_0(1710)$ shows 
the inverse pattern. This has been attributed to strong mixing between the 
pure glue basis state and the scalar quarkonium 
$|n\bar n\rangle$
and
$|s\bar s\rangle$
basis states, analogous to $\eta$-$\eta'$ mixing in the $0^{-+}$ sector
(see for example the work of Amsler and Close \cite{Ams96}).
  
\vskip 0.2cm
An alternative explanation for the violation of flavor-singlet decay symmetry
in the scalar glueball candidates has been suggested by 
Sexton {\it et al.} \cite{Sex95a,Sex95b}, 
based on a LGT study of the glueball-Ps-Ps three-point function.
In this early lattice study they found a strong dependence of this 
glueball decay coupling on the mass assumed for the final pseudoscalar mesons,
which if correct would skew mixing angle determinations using 
flavor-symmetric couplings, as assumed by Amsler and Close. 
The Sexton {\it et al.} LGT couplings favored the $f_0(1710)$ over the 
$f_0(1500)$ as a scalar glueball candidate.

\vskip 0.2cm
The very important topic of glueball decay amplitudes merits more careful
consideration in future LGT studies. The decay couplings the scalar glueball
to vector meson pairs (which may be important for the $f_0(1500)$) and the 
couplings of the lightest $0^{-+}$ and $2^{++}$ excited glueballs (as regards
favored modes and total widths) are also important topics for future lattice
studies.

\section*{Studies of the underlying QCD decay mechanisms}

Strong decay amplitudes are crucial properties of hadrons,
which if understood even at a phenomenological level can be used to identify
plausible candidates for the various conventional quark model hadrons as well
as the exotica predicted by model studies and lattice QCD.

\vskip 0.2cm
Of course there is a deeper question, which is the problem of what 
fundamental interaction in QCD is responsible for hadron strong decays
at the quark-gluon level. Although we can develop phenomenological models such
as the \tpo model without understanding the decay mechanism, these models are
simply approximate descriptions of decays with unknown and perhaps large 
systematic approximations. The QCD mechanism that drives the 
open-flavor strong decays we have discussed here is not well established, 
and has been studied in surprisingly few reference. Here we will discuss the
results of two of these references, which reach rather similar 
general conclusions.   

\vskip 0.2cm
The original Cornell charmonium group perhaps surprisingly did not assume 
the \tpo decay model. Instead they assumed a nonperturbative microscopic 
model for strong decays, in which $q\bar q$ pair production took place 
through the linear confining interaction, treated as an exchange between the 
constituent $c$ and $\bar c$ quarks and the produced light $q\bar q$ pair
(see Fig.2). This model is interesting in part because it is so highly 
constrained; the numerical value of the string tension is reasonably well
known from spectroscopy, and therefore gives decay predictions that have
no free parameters. This model gives rather good predictions for the 
behavior of $R$ (specifically charmonium production above open charm 
threshold), which implicitly involves width calculations for the higher
vector resonances. (These resonances were not treated individually 
in the Cornell studies, instead the contributions to $R$ were determined
implicitly using an effecting interaction containing decay loops.)
Nonetheless the assumption of a vector confining interaction is 
controversial, and would presumably fail the D/S ratio test in decays
such as $b_1\to \omega \pi$.

\vskip 0.2cm
A more recent paper by Ackleh {\it et al.} \cite{Ack96} 
investigates various possible QCD mechanisms for strong decays.
This reference concludes that the naive OGE pair production diagram 
in most cases gives a rather small contribution to the decay amplitude,
which (as was assumed by the Cornell group) is instead dominated 
by pair production from the linear confining interaction. Unlike the 
Cornell model however, Ackleh {\it et al.} assume a scalar confining
interaction, which gives a D/S ratio for $b_1\to \omega \pi$ that is close 
to experiment. Again the decay rates are known in this type of model
in terms of the string tension; the simple linear scalar potential
gives somewhat larger decay amplitudes than are observed 
experimentally.

\section*{A Novel Application: V-Ps Scattering from FSI}

Watson's theorem implies that final state interactions induce a phase
factor of $e^{i\delta_f}$ in the decay amplitude of a resonance into
the channel $f$. If there is only a single distinguishable channel, as in
$\rho \to \pi\pi$, this phase is not observable. However in many decays
a given final state spans several different channels, which are 
distinguished for example by internal angular momenta.

\vskip 0.2cm
An example of this FSI effect was recently exploited by 
Nozar {\it et al.} \cite{Noz02} in a very 
interesting new ``application" of strong decays. The  
decay $b_1 \to \omega \pi$ is well known as a textbook example of a meson
decay to more than one channel, since the $\omega \pi$ system can be in both
S-wave ($^3$S$_1$) and D-wave ($^3$D$_1$) states. The ratio of D-wave to
S-wave amplitudes was very important historically in selecting the \tpo model
as a realistic description of pair production quantum numbers in meson decays.

\vskip 0.2cm
Since the S- and D-wave states have different scattering 
phase shifts, they will have different FSI phases; after rescattering the 
final $\omega \pi$ state will be of the form

\be
|\omega \pi\, \rangle =
a_S\, e^{i\delta_S}\, |\omega\pi ({}^3{\rm S}_1)\rangle
+
a_D\, e^{i\delta_D}\, |\omega\pi ({}^3{\rm D}_1)\rangle \ ,
\ee  
so the famous D/S ratio is actually a complex number. The relative phase
$\delta_S - \delta_D$ appears in the $a_D a_S$ cross term in the 
squared amplitude,
and thus can be inferred directly
from the observed $\omega \pi$ angular distribution. 
The study of $b_1$ decay can thus give us very exciting information on the 
scattering amplitudes of strongly unstable resonances, which would not 
otherwise be accessible. 

\vskip 0.2cm
This measurement was carried out by 
Nozar {\it et al.} \cite{Noz02}, who found an $\omega \pi$ 
relative scattering phase at the $b_1$ mass of 
\be
\bigg[ \delta_S - \delta_D \bigg]_{\omega \pi} 
= -10.54^{\dego} \pm 2.4^{\dego} \ .
\ee 

\begin{figure}[t]
\resizebox{16pc}{!}{\includegraphics{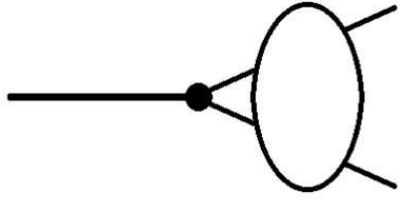}}
\caption{The meson decay FSI effect exploited by
Nozar {\it et al.} \cite{Noz02} to study $\omega\pi $ scattering.}
\end{figure}

\vskip 0.2cm
This is numerically rather similar to a quark model prediction
of $\delta_S(\omega \pi) = -14^{\dego} $ by Barnes, Black and Swanson
\cite{Bar01}. 
(One expects the $\omega \pi$ D-wave phase shift to be small at the
$b_1$ mass.)

\vskip 0.2cm
Clearly this is just one example of a large number of scattering amplitudes
of strongly unstable resonances that can be inferred from high-statistics
studies of strong decays, and future measurements should provide 
phase shifts that will be very interesting for theorists attempting to 
understand hadron scattering and FSI effects. 

\vskip 0.2cm
A caution is appropriate; we have assumed that the underlying decay
amplitudes themselves are relatively real in different channels, and also
that FSI effects are diagonal. Of course both these assumptions are suspect,
and should be tested by comparing scattering phase shifts 
inferred from the decays of different initial mesons.

\section*{Future Decays: Unquenching the Quark Model}

One of the exciting new discoveries \cite{Aub03}
discussed at this meeting was the
observation of the charm-strange mesons 
D$_{sJ}^*(2317)$ 
and 
D$_{sJ}^*(2357)$, which are presumably $0^+$ and $1^+$ states respectively.
The masses of these states are quite surprising, since the normally
accurate potential model of Godfrey and Isgur \cite{God85}
predicts much higher masses
of 2.48~GeV and 2.55~GeV respectively.

\vskip 0.2cm
Since these states are predicted by the \tpo model to be very broad
\cite{God91} and are quite close to DK threshold, one possibility is that 
the usual neglect of decay couplings in quark model calculations is inaccurate
here, and the states have been displaced downwards by ca. 150~MeV due to
decay loops. This may imply that the states have 
large DK and DK$^*$ molecular
components respectively \cite{Bar03b}, 
rather like the K$\bar {\rm K}$ molecules, 
instead of being simple $c\bar s$ quark model states.

\vskip 0.2cm
This topic of the contribution of virtual decay loop diagrams
to hadron properties,
known as ``unquenching the quark model", is an important but 
rather obscure issue. 
Explicit evaluation typically finds that individual loop 
diagrams are large, but that there may be significant cancellations
(see for example \cite{Gei91}). Future studies of strong decays will 
undoubtedly include investigations of the effects of these virtual decay
loops, since the D$_{sJ}^*(2317)$ and D$_{sJ}^*(2357)$ may have ``announced"
to us that in some strongly-coupled channels these effects cannot be ignored.

\begin{figure}[ht]
  \resizebox{16pc}{!}{\includegraphics{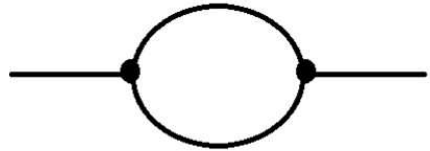}}
\caption{The generic loop diagram representing the second-order 
effects of virtual decay of a meson into 
the two-meson continuum.}
\end{figure}
                                                                                     

\begin{theacknowledgments}
I would like to thank the organizers of HADRON03 for their kind invitation 
to present this summary of the history, current status and possible future 
of research in hadron strong decays.
This research was supported in part by the U.S. National Science 
Foundation through grant NSF-PHY-0244786 
at the University of Tennessee, and by
the U.S. Department of Energy under contract
DE-AC05-00OR22725 at
Oak Ridge National Laboratory (ORNL).

\end{theacknowledgments}

\doingARLO[\bibliographystyle{aipproc}]
          {\ifthenelse{\equal{\AIPcitestyleselect}{num}}
             {\bibliographystyle{arlonum}}
             {\bibliographystyle{arlobib}}
          }
\bibliography{barnes}

\hyphenation{Post-Script Sprin-ger}
\begin{thebibliography}{33}
\expandafter\ifx\csname natexlab\endcsname\relax\def\natexlab#1{#1}\fi
\providecommand{\enquote}[1]{``#1''}
\expandafter\ifx\csname url\endcsname\relax
  \def\url#1{\texttt{#1}}\fi
\expandafter\ifx\csname urlprefix\endcsname\relax\def\urlprefix{URL }\fi

\bibitem[L.Micu(1969)]{Micu69}
L.Micu, \emph{Nucl. Phys. B}, \textbf{10}, 521 (1969).

\bibitem[{A.LeYaouanc, L.Oliver, O.P\'ene and J.C.Raynal}(1973)]{LeY73}
{A.LeYaouanc, L.Oliver, O.P\'ene and J.C.Raynal}, \emph{Phys. Rev. D},
  \textbf{8}, 2223 (1973).

\bibitem[{\it ibid.}(1974)]{LeY74}
{\it ibid.}, \emph{Phys. Rev. D}, \textbf{9}, 1415 (1974).

\bibitem[{\it ibid.}(1975)]{LeY75}
{\it ibid.}, \emph{Phys. Rev. D}, \textbf{11}, 1272 (1975).

\bibitem[{\it ibid.}(1977)]{LeY77}
{\it ibid.}, \emph{Phys. Lett. B}, \textbf{71}, 397 (1977).

\bibitem[{E.S.Ackleh, T.Barnes and E.S.Swanson}(1996)]{Ack96}
{E.S.Ackleh, T.Barnes and E.S.Swanson}, \emph{Phys. Rev. D}, \textbf{54}, 6811
  (1996).

\bibitem[{E.Eichten, K.Gottfried, T.Kinoshita, K.D.Lane and
  T.-M.Yan}(1976)]{Eic76}
{E.Eichten, K.Gottfried, T.Kinoshita, K.D.Lane and T.-M.Yan}, \emph{Phys. Rev.
  Lett.}, \textbf{36}, 500 (1976).

\bibitem[{\it ibid.}(1978)]{Eic78}
{\it ibid.}, \emph{Phys. Rev. D}, \textbf{17}, 3090 (1978).

\bibitem[{\it ibid.}(1980)]{Eic80}
{\it ibid.}, \emph{Phys. Rev. D}, \textbf{21}, 203 (1980).

\bibitem[{E.S.Swanson}(1998)]{Swa98}
{E.S.Swanson}, \emph{Nucl. Phys. Proc. Suppl.}, \textbf{64}, 312 (1998).

\bibitem[{R.M.Baltrusaitis et al.}(1986)]{Bal86}
{R.M.Baltrusaitis et al.}, \emph{Phys. Rev. Lett.}, \textbf{56}, 107 (1986).

\bibitem[{S.Godfrey, R.Kokoski and N.Isgur}(1984)]{God84}
{S.Godfrey, R.Kokoski and N.Isgur}, \emph{Phys. Lett. B}, \textbf{141}, 439
  (1984).

\bibitem[{H.G.Blundell and S.Godfrey}(1996)]{Blu96}
{H.G.Blundell and S.Godfrey}, \emph{Phys. Rev. D}, \textbf{53}, 3700 (1996).

\bibitem[{S.Godfrey and N.Isgur}(1985)]{God85}
{S.Godfrey and N.Isgur}, \emph{Phys. Rev. D}, \textbf{32}, 189 (1985).

\bibitem[{S.Capstick and N.Isgur}(1986)]{Cap86}
{S.Capstick and N.Isgur}, \emph{Phys. Rev. D}, \textbf{34}, 2809 (1986).

\bibitem[{T.Barnes, F.E.Close, P.R.Page and E.S.Swanson}(1997)]{Bar97}
{T.Barnes, F.E.Close, P.R.Page and E.S.Swanson}, \emph{Phys. Rev. D},
  \textbf{55}, 4157 (1997).

\bibitem[{T.Barnes, N.Black and P.R.Page}(2003)]{Bar03a}
{T.Barnes, N.Black and P.R.Page}, \emph{Phys. Rev. D}, \textbf{68}, 054014
  (2003).

\bibitem[{S.Capstick and W.Roberts}(1994)]{Cap94}
{S.Capstick and W.Roberts}, \emph{Phys. Rev. D}, \textbf{49}, 4570 (1994).

\bibitem[{\it ibid.}(2000)]{Cap00}
{\it ibid.}, \emph{Prog. Part. Nucl. Phys.}, \textbf{45}, {S241} (2000).

\bibitem[{H.J.Lipkin}(2000)]{Lip00}
{H.J.Lipkin}, \emph{Phys. Lett. B}, \textbf{494}, 248 (2000).

\bibitem[{N.Isgur, R.Kokoski and J.Paton}(1985)]{Isg85}
{N.Isgur, R.Kokoski and J.Paton}, \emph{Phys. Rev. Lett.}, \textbf{54}, 869
  (1985).

\bibitem[{F.E.Close and P.R.Page}(1995)]{Clo95}
{F.E.Close and P.R.Page}, \emph{Nucl. Phys. B}, \textbf{443}, 233 (1995).

\bibitem[{C.McNeile, C.Michael and P.Pennanen (UKQCD
  Collaboration)}(2002)]{McN02}
{C.McNeile, C.Michael and P.Pennanen (UKQCD Collaboration)}, \emph{Phys. Rev.
  D}, \textbf{65}, 094505 (2002).

\bibitem[{C.J.Morningstar and M.Peardon}(1999)]{Mor99}
{C.J.Morningstar and M.Peardon}, \emph{Phys. Rev. D}, \textbf{60}, 034509
  (1999).

\bibitem[{C.Amsler and F.E.Close}(1996)]{Ams96}
{C.Amsler and F.E.Close}, \emph{Phys. Rev. D}, \textbf{53}, 295 (1996).

\bibitem[{J.Sexton, A.Vaccarino and D.Weingarten}(1995{\natexlab{a}})]{Sex95a}
{J.Sexton, A.Vaccarino and D.Weingarten}, \emph{Nucl. Phys. B}, \textbf{42},
  279 (1995{\natexlab{a}}).

\bibitem[{J.Sexton, A.Vaccarino and D.Weingarten}(1995{\natexlab{b}})]{Sex95b}
{J.Sexton, A.Vaccarino and D.Weingarten}, \emph{Phys. Rev. Lett.}, \textbf{75},
  4563 (1995{\natexlab{b}}).

\bibitem[{M.Nozar {\it et al.} (E852 Collaboration)}(2002)]{Noz02}
{M.Nozar {\it et al.} (E852 Collaboration)}, \emph{Phys. Lett. B},
  \textbf{541}, 35 (2002).

\bibitem[{T.Barnes, N.Black and E.S.Swanson}(2001)]{Bar01}
{T.Barnes, N.Black and E.S.Swanson}, \emph{Phys. Rev. C}, \textbf{63}, 025204
  (2001).

\bibitem[{B.Aubert {\it et al.} (BABAR Collaboration)}(2003)]{Aub03}
{B.Aubert {\it et al.} (BABAR Collaboration)}, \emph{Phys. Rev. Lett.},
  \textbf{90}, 242001 (2003).

\bibitem[{S.Godfrey and R.Kokoski}(1991)]{God91}
{S.Godfrey and R.Kokoski}, \emph{Phys. Rev. D}, \textbf{43}, 1679 (1991).

\bibitem[{T.Barnes, F.E.Close and H.J.Lipkin}(2003)]{Bar03b}
{T.Barnes, F.E.Close and H.J.Lipkin}, \emph{Phys. Rev. D}, \textbf{68}, 054006
  (2003).

\bibitem[{P.Geiger and N.Isgur}(1991)]{Gei91}
{P.Geiger and N.Isgur}, \emph{Phys. Rev. Lett.}, \textbf{67}, 1066 (1991).

\end{thebibliography}

\end{document}